**Title:**

# ASCHOPLEX encounters Dafne: a federated continuous learning project for the generalizability of the Choroid Plexus automatic segmentation

**Authors:**
Valentina Visani[1,2], Marco Pinamonti[1], Valentina Sammassimo[1], Manuela Moretto[1,3], Mattia Veronese[1,3], Agnese Tamanti[4], Francesca Benedetta Pizzini[5], Massimiliano Calabrese[4], Marco Castellaro[1], and Francesco Santini[2,6]

[1] *Department of Information Engineering, University of Padova, Padova, Italy*
[2] *Basel Muscle MRI, Department of Biomedical Engineering, University of Basel, Basel, Switzerland*
[3] *Department of Neuroimaging, Institute of Psychiatry, Psychology and Neuroscience, King's College London, London, UK*
[4] *Department of Neurosciences, Biomedicine and Movement Sciences, University of Verona, Verona, Italy*
[5] *Department of Engineering for Innovation Medicine, University of Verona, Verona, Italy*
[6] *Department of Radiology, University Hospital Basel, Basel, Switzerland*

**Abstract:**
The Choroid Plexus (ChP) is a highly vascularized brain structure that plays a critical role in several physiological processes. ASCHOPLEX, a deep learning–based segmentation toolbox with an integrated fine-tuning stage, provides accurate ChP delineations on non–contrast-enhanced T1-weighted MRI scans; however, its performance is hindered by inter-dataset variability. This study introduces the first federated incremental learning approach for automated ChP segmentation from 3D T1-weighted brain MRI, by integrating an enhanced version of ASCHOPLEX within the Dafne (Deep Anatomical Federated Network) framework. A comparative evaluation is conducted to assess whether federated incremental learning through Dafne improves model generalizability across heterogeneous imaging conditions, relative to the conventional fine-tuning strategy employed by standalone ASCHOPLEX. The experimental cohort comprises 2,284 subjects, including individuals with Multiple Sclerosis as well as healthy controls, collected from five independent MRI datasets. Results indicate that the fine-tuning strategy provides high performance on homogeneous data (e.g., same MRI sequence, same cohort of subjects), but limited generalizability when the data variability is high (e.g., multiple MRI sequences, multiple and new cohorts of subjects). By contrast, the federated incremental learning variant of ASCHOPLEX constitutes a robust alternative consistently achieving higher generalizability and more stable performance across diverse acquisition settings.

**Introduction:**

*The role of the Choroid Plexus*
The choroid plexus (ChP) is a highly vascularized structure located within the four cerebral ventricles, primarily responsible for the production of cerebrospinal fluid (Damkier et al., 2013) and for maintaining brain homeostasis and immune surveillance (Balusu et al., 2016; Scarpetta et al., 2025; Spector et al., 2015). Recent studies have further emphasized the ChP essential role in modulating the brain's immune response (Courtney et al., 2025), positioning it as a key interface between the central nervous system and peripheral immunity. Moreover, a growing body of evidence suggests that an increase in choroid plexus volume (ChPV) may serve as a non-invasive marker of neuroinflammation in neurodegenerative disorders, including Multiple Sclerosis (MS) (De Meo et al., 2025; Engelhardt et al., 2001; Fleischer et al., 2021; Kolahi et al., 2024; Lassmann, 2019; Monaco et al., 2020; Ricigliano et al., 2021; Vercellino et al., 2008), Parkinson's disease and Alzheimer's disease (Tadayon et al., 2020a, 2020b), as well as in psychiatric conditions (Althubaity et al., 2022; Zhou et al., 2020). However, the relationship between ChPV and underlying inflammatory processes remains complex and multifactorial (Magliozzi et al., 2025). Even if the first applications of ChP imaging concerned pathological conditions (Bergsland et al., 2024), in the recent years few studies are starting focusing on small healthy cohorts (Alisch et al., 2022, 2021; Eisma et al., 2024, 2023), aiming to investigate the relationship between ChPV and aging.

*Current approaches to Choroid Plexus Segmentation*
Quantitative assessment of the ChPV from structural Magnetic Resonance Imaging (MRI) necessitates precise anatomical segmentation to deeply understand its pathophysiological role (Alisch et al., 2021; Althubaity et al., 2022; Duan and Qi, 2023; Eisma et al., 2023). Gadolinium enhanced T1-weighted (T1-w) MRI is the gold standard sequence for imaging the ChP (Hubert et al., 2019; Tadayon et al., 2020a). However, non-contrast-enhanced T1-w sequences are often preferred in routine clinical practice due to their non-invasive nature and because they have been shown to provide ChPV estimates comparable to those obtained with Gadolinium-enhanced sequences (Senay et al., 2023; Visani et al., 2024a). Manual segmentation is often referred as the Ground Truth (GT) for ChPV analysis but is inherently limited by its time-consuming nature, operator dependency, and both the anatomical complexity and low tissue contrast of the ChP (Tadayon et al., 2020a; Yazdan-Panah et al., 2023). These limitations have encouraged the development of automated segmentation techniques. Among them, deep learning-based approaches (Eisma et al., 2024; Visani et al., 2024b; Yazdan-Panah et al., 2023), mostly based on Deep Neural Networks (DNN), have demonstrated superior scalability and consistency when compared to traditional tools such as FreeSurfer (Fischl, 2012) and its Gaussian Mixture Model-enhanced versions (Storelli et al., 2024; Tadayon et al., 2020a). Recent advances, such as the ASCHOPLEX toolbox (Visani et al., 2024b), have achieved state-of-the-art segmentation accuracy by leveraging architectures including nnUNet (Isensee et al., 2021) and UNETR (Hatamizadeh et al., 2022b), implemented through majority-voting ensembles and fine-tuned for domain-specific imaging protocols. Although fine-tuning can enhance model performance on new datasets, its effectiveness depends heavily on the availability of manually annotated labels (Yan et al., 2020). Moreover, when data are collected from multiple sources, labeled examples are required for each scanner type or acquisition setting. Furthermore, data protection regulations such as HIPAA (*HIPAA - US Department of Health and Human Services*, 2020) and GDPR (*GDPR - General Data Protection Regulation*, 2019) restrict the sharing of annotated data across institutions, complicating the development of diverse and representative centralized training sets (Raudys and Jain, 1991). These factors hinder centralized data collection and model training, particularly in multicenter medical studies where data heterogeneity presents an additional challenge.

*Federated learning for improving generalizability*

Federated learning has emerged as a promising strategy to overcome these constraints (Wen et al., 2023). In fact, it enables collaborative model training across multiple institutions without requiring data to leave local premises. Instead, only model updates (e.g., gradients or weights) are shared and aggregated to form a global model, preserving data confidentiality and complying with legal and ethical standards (Adnan et al., 2022; Guan et al., 2024). This distributed approach is particularly suitable for medical applications, where patient data are highly sensitive and regulatory compliance is mandatory (Kaissis et al., 2020). Furthermore, by aggregating knowledge from diverse imaging sources and populations, federated learning promotes model generalization across heterogeneous clinical domains.

To enhance adaptability, federated learning can be combined with incremental (or continuous) learning, a strategy that allows models to incrementally incorporate new data or tasks without overwriting previously learned information (Hamedi et al., 2025). This capability is essential in medical imaging—particularly MRI—where acquisition protocols, scanner configurations, and patient demographics might vary over time (Cai et al., 2020). This approach allows for continuous model refinement, making it possible to maintain performance levels even as new types of data and clinical conditions emerge (Nazir and Kaleem, 2023). The integration of federated and incremental learning in medical imaging has been explored in several studies. However, for image segmentation tasks, incorporating such models into clinical workflows requires not only high accuracy, but also expert oversight enabling clinicians to control over the final segmentation and ensuring clinical responsibility (Castiglioni et al., 2021; Ng et al., 2021). The necessary presence of a human verifier in the clinical workflow provides an ideal context for implementing continuous learning. With expert feedback, models can be continuously refined and adapted to new data, improving their performance over time (Shen et al., 2021). This iterative process ensures that segmentation models remain relevant and accurate, addressing the dynamic nature of medical imaging data and clinical requirements (Yi et al., 2020). A possibility is using image visualization interfaces, allowing the user to maintain final control over the segmentation accuracy (Shen et al., 2021).

A practical and efficient solution is represented by Dafne (Deep Anatomical Federated Network) (Santini et al., 2025), an open-source, Python-based client-server platform specifically designed for federated incremental learning in the medical image segmentation domain. Dafne operates by distributing a global model from a centralized, secure cloud server to decentralized client nodes, where training is performed incrementally on small, local datasets. Updated models are subsequently aggregated into the global model without discarding prior knowledge, thereby mitigating catastrophic forgetting (Aleixo et al., 2023). Dafne also incorporates an expert validation interface and has already been successfully deployed in segmenting various anatomical regions based on 2D DNN models. These features make it a compelling candidate for developing adaptive, privacy-compliant, and clinically robust segmentation frameworks.

*The aim of this work*

The aim of this study is to develop the first federated incremental learning framework for the automated segmentation of the ChP from 3D T1-w brain MRI. To this end, we integrated the ASCHOPLEX pipeline into the Dafne platform. Concurrently, the ASCHOPLEX framework was modified with the addition of SwinUNETR, a state-of-the-art transformer-based segmentation model, alongside its existing UNETR and DynUNet backbones. The proposed method was evaluated using a clinical validation sample including five independent MRI datasets comprising both MS patients and healthy controls (HC). Through comparative analysis simulating a multi-center study, we hypothesized that federated incremental learning via Dafne yields superior generalizability across heterogeneous imaging conditions (e.g., multiple MRI sequences, multiple cohorts, presence of artifacts), relative to the conventional fine-tuning approach employed by standalone ASCHOPLEX.

**Materials & Methods:**

The overall workflow of the study is illustrated in **Figure 1**. T1-w MRI data from five selected datasets (see *Datasets*) were first subjected to quality control procedures (see *Quality Check of T1-weighted MRI*) and subsequently manually segmented. The starting ASCHOPLEX model (Model 0) was then obtained through a five-fold cross-validation training procedure on the first dataset (see *ASCHOPLEX initial training*). The resulting pre-trained model was then improved using two different strategies on the other datasets. In the first strategy, separate fine-tuning (FT) steps were performed independently on each dataset. In the second strategy, implemented in Dafne, an incremental learning (IL) approach was adopted, whereby datasets were incorporated sequentially. The performance of these two approaches were then compared.

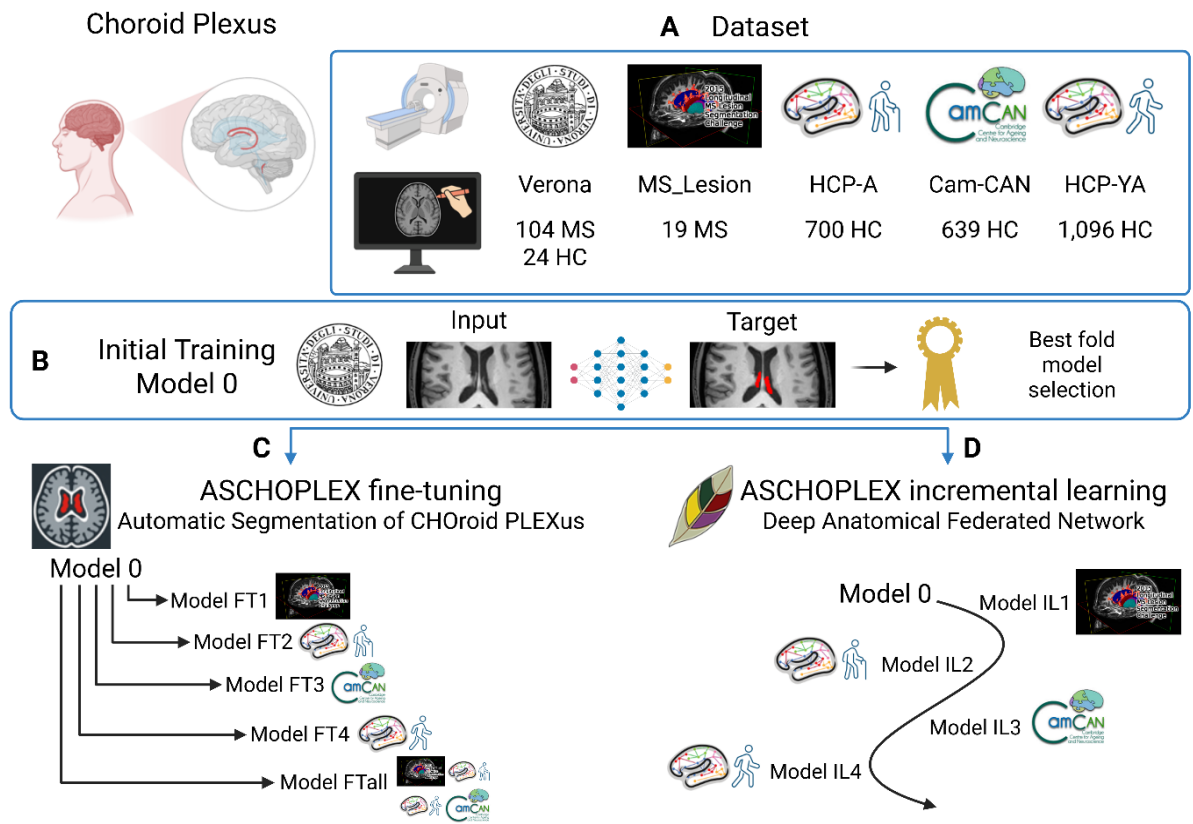

**Figure 1:** *Workflow of the study design. (A)T1-w MRI data derived from five datasets were manually segmented (HC: Healthy Controls, MS: Multiple Sclerosis; Verona: Multiple Sclerosis Centre of the University of Verona; MS_Lesion: ISBI 2015 Longitudinal MS Lesion Segmentation Challenge; HCP-A: Lifespan 2.0 Release of Human Connectome Project in Aging; Cam-CAN: Cambridge Centre for Ageing and Neuroscience; HCP-YA: Human Connectome Project Young Adult). (B) ASCHOPLEX initial five-fold cross-validation training on Verona dataset (Model 0), with the selection of the best model for each fold based on Dice Coefficient and the ensemble by major voting of the five selected model predictions to obtain the final outcome of ASCHOPLEX. (C) Separate ASCHOPLEX fine-tuning of the pre-trained Model 0 on other datasets (Model FT1-4, FTall). (D) ASCHOPLEX incremental learning with Dafne, starting from Model 0 and sequentially incorporating other datasets (Model IL1-4). Illustration created with BioRender.com.*

*Datasets*

A total of 2,582 subjects from five datasets were initially included in this study, all of which contained multiple MRI sequences, although only the T1-w MRI sequence was considered for the purposes of this work. The Verona dataset was provided by the Multiple Sclerosis Centre of the University Hospital of Verona, and it was composed of 128 MS subjects derived from two different scanners. The MS_Lesion dataset included the first time point for each of the 19 MS subjects of the ISBI 2015 Longitudinal MS Lesion Segmentation Challenge (Carass et al., 2017a, 2017b). The HCP-A dataset was composed by a subset of 700 HC subjects derived from the Lifespan 2.0 Release of Human Connectome Project in Aging (Bookheimer et al., 2019). The Cam-CAN dataset considered 639 HC subjects of the public dataset released by the Cambridge Centre for Ageing and Neuroscience (Shafto et al., 2014). Finally, the HCP-YA dataset comprised 1,096 HC subjects released by the public dataset of the Human Connectome Project Young Adult (HCP-YA) (Van Essen et al., 2013). A detailed description of the datasets is reported in **Table *1***.

The subjects inside each dataset were divided into a training set and a testing set (**Table *1***). The images from the Verona dataset were used for the initial training; therefore, the first training set consisted of 92 subjects, while remaining 36 subjects composed the testing set. This dataset was chosen as the starting one to ensure consistency with the first version of ASCHOPLEX, which was trained on the same data. The other four datasets were used for the fine-tuning and the incremental learning (with Dafne) procedures of ASCHOPLEX; consequently only 10 subjects for each of these datasets were assigned to the training set of both incremental learning and fine-tuning steps. Data splitting was performed in blind fashion, and for each dataset we maintained a balanced distribution of controls and patients, males and females, younger and elder subjects, and scanner type.

To provide a more comprehensive assessment of each model's performance, we constructed a composite dataset, hereafter referred to as General. For each evaluated model, this dataset includes 10 subjects from each individual dataset (9 for MS_Lesion), selected to ensure that their Dice coefficients are approximately evenly distributed around the median. This results in a total of 49 subjects per model. By using this sampling strategy, we aim to obtain a robust representation of model performance that is less affected by outliers inherent to the individual datasets.

**Table 1:** *Subjects information and MRI acquisition protocol for the five datasets. Subjects' division into training and testing sets for both the initial training and the subsequent fine-tuning and incremental learning approaches.*

| Dataset | Subjects Details | Acquisition Protocol | Subjects' division |
|---|---|---|---|
| | | *Initial training* | |
| Verona | 67 sj<br>M/F 21/46<br>24 HC<br>age 37.2±9.5 y<br>43 MS<br>age 40.9±9.9 y | 3T Philips Achieva TX<br>8-channels head coil (R3.2.3.2)<br>3D T1-weighted MPRAGE<br>resolution: 1x1x1 mm$^3$<br>FA: 9°; SENSE: 2.5<br>TE/TR/TI: 3.7/8.4/1037 ms<br>Acquisition time: 4 min 50 s | Training: 29 MS, 17 HC<br>Testing: 14 MS, 7 HC |
| | 61 MS<br>M/F 13/48<br>age 36.7±10.1 y | 3T Philips Elition S<br>32-channels head coil (R5.7.2.1)<br>3D T1-weighted MPRAGE<br>resolution: 1x1x1 mm$^3$<br>FA: 8°; SENSE: 4<br>TE/TR/TI: 3.7/8.4/1016 ms<br>Acquisition time: 3 min 20 s | Training: 46 MS<br>Testing: 15 MS |
| | Total: 128 sj | | Training: 92 sj<br>Testing: 36 sj |
| | | *Fine-tuning and Incremental learning* | |
| MS_Lesion | 19 MS<br>M/F 4/15<br>age 43.5±10.3 y | 3T Philips<br>3D T1-weighted MPRAGE<br>resolution: 0.82x0.82x1.17 mm$^3$; FA: 8°;<br>TE/TR: 6/10.3 ms | Training: 10 sj<br>Testing: 9 sj |
| HCP-A | 700 HC<br>M/F 309/391<br>age 36-88 y | 3T Siemens MAGNETOM Prisma<br>32-channels head coil<br>3D T1-w multi-echo MPRAGE<br>resolution: 0.8x0.8x0.8 mm³; FA: 8°;<br>TEs: 1.8, 3.6, 5.4, 7.2 ms;<br>TR/TI: 2500/1000 ms<br>Total acquisition time: 7 min 40 sec.<br>The average of the first two echoes was employed for further analyses. | Training: 10 sj<br>Testing: remaining sj |
| Cam-CAN | 639 HC<br>M/F 315/324<br>age 22-88 y | 3T Siemens TIM Trio System<br>32-channels head coil<br>3D T1-weighted MPRAGE<br>resolution: 1x1x1mm³; FA: 9°;<br>TE/TR/TI: 2.99/2250/900 ms<br>Acquisition time: 4 min 32 sec | Training: 10 sj<br>Testing: remaining sj |
| HCP-YA | 1,096 HC<br>M/F 500/596<br>age 22-35 y | 3T Siemens Connectome Skyra<br>32-channel head coil<br>3D T1-weighted MPRAGE<br>resolution: 0.7x0.7x0.7 mm³; FA: 8°;<br>TE/TR/TI: 2.14/2500/1000 ms<br>Acquisition time: 7 min 40 sec | Training: 10 sj<br>Testing: remaining sj |

*Legend: y = years, sj = subjects, MS = Multiple Sclerosis, HC = Healthy Controls, Verona = Multiple Sclerosis Centre of the University of Verona, MS_Lesion = ISBI 2015 Longitudinal MS Lesion Segmentation Challenge, HCP-A= Lifespan 2.0 Release of Human Connectome Project in Aging, Cam-CAN = Cambridge Centre for Ageing and Neuroscience, HCP-YA = Human Connectome Project Young Adult.*

*Quality-check of T1-w MRI*

Prior to further processing, the structural MRI data from the Cam-CAN, HCP-A, and HCP-YA cohorts underwent an initial quality screening using the MRIQC software version 24.1.0 (Esteban et al., 2017). This tool quantitatively evaluated scan quality through a set of standardized metrics, encompassing noise characteristics, information theory, and indicators of common imaging artifacts. These quantitative outputs were plotted for all subjects within each dataset to facilitate the detection of atypical scans. Imaging data that deviated substantially from the normative range for any of these metrics underwent additional visual inspection and were excluded when warranted. Concerning Verona and MS_Lesion datasets, the quality check was conducted by visual inspection [V.V.].

*ChP manual segmentation*
The GT segmentation was generated in the native T1-w space and restricted to the two lateral ventricles, as on conventional 3T MRI the ChP can be barely visualized in the third and fourth ventricles and therefore it could not be segmented with sufficient reliability (Senay et al., 2023; Visani et al., 2024b). GTs were manually depicted using ITK-SNAP (Yushkevich et al., 2006). For the Verona dataset, the GT was performed for all the 128 subjects by a senior neuroradiologist [F.B.P.]. For the MS_Lesion dataset, the GT was performed for all the 19 subjects by a trained researcher [V.V.]. Concerning HCP-A, Cam-CAN, and HCP-YA datasets, the GT for the ten training subjects was performed manually by a trained researcher [V.V.]. For all the other subjects, the first version of ASCHOPLEX (Visani et al., 2024b) was separately fine-tuned on the ten manually labelled subjects for each of the three datasets and the output segmentations were manually checked and corrected by a trained researcher [V.V.]; the checked segmentations were used as GT. The resulting GT served as a reference for evaluating both the fine-tuning and incremental learning methods.

*ASCHOPLEX initial training*
The original version of ASCHOPLEX was implemented in MONAI v 1.0.1 (MONAI Consortium, 2022) and the released version considered an ensemble by major voting of UNETR (Hatamizadeh et al., 2022b) and DynUNet, the MONAI version of nnU-Net (Isensee et al., 2021) architectures, each architecture used 128 x 128 x 128 pixels isotropic patches with Generalized Dice or combination of both Dice and cross-entropy loss functions. For the purpose of this study, ASCHOPLEX version was updated as follows. First, the MONAI version employed was 1.4.0 (MONAI Consortium, 2024), which was up to date at the time of the analyses conducted in this study. Second, the pool of tested architectures was expanded to include, in addition to the previously selected UNETR and DynUNet, the SwinUNETR architecture (Hatamizadeh et al., 2022a). Dice and cross-entropy loss function configurations were considered based on previous findings (Visani et al., 2024b).

The Verona dataset was used for the initial training of the networks, balancing HC/MS and the scanner type in the training/validation/testing steps (**Table *1***). The testing set was left untouched for the comparison. The DynUnet, a self-configuring DNN, was configured as in the original version of ASCHOPLEX. The model processed 3D images with a single input channel and was structured into four main blocks: input, downsampling, bottleneck, and upsampling. Downsampling and upsampling operations were performed using convolutions with strides of [1, 2, 2, 1] and a kernel size of $3 \times 3 \times 3$ voxels. Each convolution was followed by instance normalization and a LeakyReLU activation function. UNETR incorporated a ViT-based encoder connected to a convolutional decoder via skip connections, effectively combining the ability of transformers to capture global contextual information with the U-Net's strength in preserving spatially localized features. The architecture was configured with a feature size of 16, a hidden layer dimension of 768, and a feedforward network size of 3072. The transformer encoder employed 12 self-attention heads. To retain spatial information across input tokens, a learnable positional embedding based on a multi-layer perceptron was utilized. At each resolution level, the token embeddings were reshaped and projected back to the spatial domain via a sequence of $3 \times 3 \times 3$ convolutional layers, each followed by instance normalization. Differently from the UNETR, Swin-UNETR integrated a Swin Transformer encoder with a U-Net–based convolutional decoder, effectively combining global context modeling with fine-grained spatial localization. The encoder partitioned the 3D input volume into non-overlapping patches and applied self-attention within local windows, as defined by a patch partitioning mechanism. This hierarchical design enabled efficient long-range dependency modeling while preserving computational efficiency. Feature representations extracted at multiple resolutions were connected to the decoder via skip connections, facilitating multi-scale information fusion. The model was designed to process single-channel 3D inputs, with an initial feature size of 48.

As for the original version of ASCHOPLEX, the five-folds cross validation training procedure considered data augmentation transforms to extend model variability. The data augmentation operation considered rotations and flip among three axes (p=15%) and intensity shifts (p=15%) as in the first version of ASCHOPLEX, but with the addition of random bias field augmentation (p=15%) and Gibbs noise artifacts (p=15%), and the TorchIo motion artifacts (p=30%), elastic deformations (p=30%), and ghost artifacts (p=30%). The image intensity was normalized between 0 and 1. Each configuration was trained using the Adam-Weighted optimizer (Loshchilov and Hutter, 2019) with 1e-04 learning rate, 1e-05 weight decay, maximum number of iterations fixed at 3e04, and single batch size. The training was performed on a 48 GB NVIDIA Tesla A40 GPU. Once the training finished, the selection of the best fold DNN configuration was made based on the validation Dice Coefficient to obtain the five best models that were then ensembled by major voting to augment the robustness of the final predictions.

*ASCHOPLEX fine-tuning*

ASCHOPLEX was released to address the deep learning-based methods limitation of model generalizability when applied on an unseen set of data. In fact, variations in scanner hardware, sequence parameters, software versions, or subject morphology can introduce untrained features, leading to performance degradation (Yan et al., 2020). In some cases, this issue can be mitigated through fine-tuning. Based on this hypothesis, ASCHOPLEX was released with the opportunity of performing a fine-tuning step on ten manually labelled subjects, where five were used for the training of the DNN and five for the validation of the model. ASCHOPLEX demonstrated its ability in learning new image features and adapting to new data thanks to the fine-tuning procedure, however, the forgetting of image characteristics from the initial training dataset was not investigated. Moreover, considering a multi-scanner dataset, the user would be requested to perform different fine-tuning steps to maximize the performance for each image type, that would affect the generalizability. This study proposed to investigate the forgetting of ASCHOPLEX when the fine-tuning procedure is applied to datasets with different characteristics. ASCHOPLEX was fine-tuned separately, after the initial training on the Verona dataset (Model 0), on the MS_Lesion (Model FT1), HCP-A (Model FT2), Cam-CAN (Model FT3), and HCP-YA (Model FT4) datasets, as well as on all four datasets combined (Model FTall), and was then inferred on the testing set of each dataset. The training settings were identical to those used for the initial training, with the exception of the maximum number of iterations, which was set to 1e04. In this phase, a single validation split was used, removing the need to partition the training set into folds. As in the previous experiments, the final prediction for each test subject was obtained through an ensemble approach, using majority voting across the segmentations produced by the five fine-tuned models. The training was performed on a 48 GB NVIDIA Tesla A40 GPU.

*ASCHOPLEX incremental learning with Dafne*

To address the limitations of classical fine-tuning—particularly those related to multi-scanner datasets and catastrophic forgetting—a federated incremental learning strategy was adopted. This approach allows the model to adapt to new data while preserving previously acquired knowledge by performing a weighted averaging of the pre-existing and the newly fine-tuned models at each aggregation step. In this context, Dafne (Santini et al., 2025) serves as a suitable platform, designed as an open-source client-server framework for computer-assisted medical image segmentation. The server, hosted on a Google Cloud Virtual Machine, distributes segmentation models to clients and integrates updated versions via a Representational state transfer Application Programming Interface. By design, Dafne enables users to obtain accurate automatic segmentations through the incremental learning step, without the need to provide manually annotated references as input to the model. In fact, users download the latest model for a selected anatomical region, apply it to the loaded medical images, and refine the segmentations through interactive tools such as spline-based contouring and nonrigid registration, thanks to a guided user interface. This enables real-time access to manual segmentation outputs by the user. After that, the client performs five epochs of local incremental learning, after which the updated model is sent back to the server. Validation is performed by calculating the Dice Coefficient on a private reference dataset, and successful updates are merged with the central model. Dafne employs a plugin-like architecture based on Python's dill serialization, supporting modular extension and various model types. Core models are based on VNet and ResNet architectures for 2D MRI segmentation, implemented using TensorFlow (Abadi et al., 2016) and Keras (Chollet, F, 2025). The models hosted in Dafne are four: six muscles of the human lower leg; twelve muscles of the thigh; the abdominal organs (liver, kidneys, spleen); the lumbar spine (vertebrae, discs, and spinal canal).

To enhance Dafne's robustness and extend its capabilities to complex multi-organ 3D segmentation, this work integrated the ASCHOPLEX toolbox into the platform. The integration required extending Dafne's compatibility with PyTorch (Paszke et al., 2019) and MONAI v1.4.0 (MONAI Consortium, 2024) libraries, and adapting the input handling to support full 3D MRI volumes rather than single slices, thus preserving spatial information, critical for accurate ChP segmentation. Moreover, the incremental learning control logic was revised: unlike the 2D models that require five segmented slices, the ASCHOPLEX integration mandates at least ten 3D volumes before initiating the local training phase, to be consistent with ASCHOPLEX original implementation. This modification enabled Dafne to dynamically differentiate between 2D and 3D workflows and, consequently, to obtain a final segmentation derived from the ensemble by major voting of the five ASCHOPLEX models. The initial model trained on the Verona dataset was the first one loaded into the Dafne platform. The incremental learning settings were the same as the ASCHOPLEX fine-tuning procedure except for the maximum number of iterations (4e03). The incremental learning was performed on one dataset at a time, following this sequential order: MS_Lesion (Model IL1), HCP-A (Model IL2), Cam-CAN (Model IL3), HCP-YA (Model IL4) dataset. For each incremental learning step, the new model was loaded on the Dafne server and the merging procedure with the previous model was performed, evaluating the Dice Coefficient on the testing set of the MS_Lesion dataset. This choice was made since this dataset is publicly available, and free for download without restriction. The incremental learning procedure was performed on an 11 GB NVIDIA GeForce GTX 1080 Ti GPU.

*Performance evaluation*

We conducted a comparative analysis between the segmentations generated by the initial ASCHOPLEX model (Model 0), ASCHOPLEX fine-tuning (Model FT1-4, Model FTall), and ASCHOPLEX incremental learning with Dafne (Model IL1-4). For the simulation of a multi-center study, a final comparison between Model FTall and Model IL4 was made to assess the generalizability of the proposed approaches. These comparisons were performed separately on the unseen testing sets of all the five datasets. The GT was used as the reference segmentation. The performance metric investigated was the Dice Coefficient. It was computed over the entire 3D volumes, excluding background voxels, to assess the similarity between predicted and reference segmentations. Optimal performance is defined as maximizing the Dice. All Dice metrics were reported as Median Interquartile Range (IQR), where the IQR was calculated as the difference between the third and the first quartile. Limited to the final comparison between Model FTall and Model IL4, the Absolute Volume [mm$^3$] of both the GT and the predicted segmentations was taken into consideration.

*Software availability*

The integration of ASCHOPLEX into Dafne proposed in this paper is freely available on GitHub (https://github.com/dafne-imaging). The updated version of ASCHOPLEX is freely available on GitLab (https://gitlab.dei.unipd.it/fair/aschoplex2.0).

**Results:**

*Quality-check of T1-w MRI*

After the MRIQC step, 567 HC (M/F 246/321, age: 36-88 years) of HCP-A, 587 HC (M/F 288/299, age: 22-88 years) of Cam-CAN, and 983 HC (M/F 446/537, age: 22-35 years) of HCP-YA were retained. Consequently, the number of subjects used as a testing set for each dataset was 557 HC for HCP-A, 577 HC for Cam-CAN, and 973 HC for HCP-YA. None of the images of both Verona and MS_Lesion were discarded. The total number of subjects employed for subsequent analyses was 2,284.

*ASCHOPLEX initial training*

The selected DNNs configurations were two SwinUNETR, two UNETR, and a DynUnet. These five configurations were used for both the fine-tuning and the incremental learning approaches.

*ASCHOPLEX fine-tuning*

**Table 2** reports the Dice Coefficient metrics on the testing subjects of all five investigated datasets comparing Model 0, FT1-4 and FTall.

What emerges is that, for each considered dataset, the model achieving the best Dice performance is the one trained or fine-tuned on that specific dataset, with the exception of HCP-YA in which the FT did not enhance the performance of the pre-trained model (Dataset: Model (FT)X - Median [IQR]): Verona: Model 0 – 0.811 [0.087]; MS_Lesion: Model FT1 – 0.695 [0.032]; HCP-A: Model FT2 - 0.826 [0.084]; Cam-CAN: Model FT3 - 0.920 [0.040]; HCP-YA: Model 0 – 0.764 [0.139]. However, as with Model 0, the fine-tuned models (FTX) perform poorly when considering the other datasets (i.e., Model FT1 on Verona: 0.580 [0.143]; Model FT3 on HCP-YA: 0.647 [0.238]). Moreover, the Model FTall generally yields lower performance compared to the Model FTX on the evaluated X dataset (i.e., Model FTall on: MS_Lesion 0.344 [0.217]; Cam-CAN 0.542 [0.252]), but taking into consideration the other datasets, its performance is moderately higher (i.e., Model FT2: HCP-A 0.826 (0.084), Cam-CAN 0.299 [0.368], Verona 0.515 [0.291]; Model FTall: HCP-A 0.789 (0.072), Cam-CAN 0.542 [0.252], Verona 0.654 [0.137]). Finally, using the General dataset as a summary of performance across all datasets, Model 0 achieves the highest median Dice value (0.765 [0.482]), followed by Model FTall (0.662 [0.215]). Although Model 0 maintains the strongest generalization on average, it fails on MS_Lesion (0.411 [0.111]) and HCP-A (0.044 [0.423]). Conversely, Model FTall brings more stable outcomes across all datasets. The other single fine-tuned models (Model FT1-4) exhibit very limited generalizability across datasets.

**Figure 2** shows the distributions of the Dice coefficients for each model across the five datasets. Consistent with the results in **Table 2**, for each model, the dataset with the lowest variability and fewest outliers is generally the one used for the training or fine-tuning - namely, Verona for Model 0, MS_Lesion for Model FT1, HCP_A for Model FT2, and Cam-CAN for Model FT3. Although the Dice distributions of Model FTall are suboptimal for most datasets when compared with those of the corresponding single fine-tuned models, the distributions on the General dataset highlight the limited generalizability of the single fine-tuned models, which exhibit higher variability and lower Dice values than Model FTall.

**Table 2:** *Dice Coefficient values on the test sets of the five datasets. The reference is the manual segmentation. Model 0 refers to the initial ASCHOPLEX trained on the Verona dataset, while Models FT1–4 correspond to the fine-tuning performed individually on MS_Lesion (Model FT1), HCP-A (Model FT2), Cam-CAN (Model FT3), and HCP-YA (Model FT4). Model FTall corresponds to the unique fine-tuning performed on all four datasets. The General dataset is composed of ten selected subjects for each of the five datasets (see Materials & Methods).*

| Dice Coefficient | Verona | MS_Lesion | HCP-A | Cam-CAN | HCP-YA | General |
|---|---|---|---|---|---|---|
| **Model 0** | **0.811 [0.087]** | 0.411 [0.111] | 0.044 [0.423] | 0.808 [0.059] | **0.764 [0.139]** | **0.765 [0.482]** |
| **Model FT1** | 0.580 [0.143] | **0.695 [0.032]** | 0.126 [0.445] | 0.636 [0.188] | 0.531 [0.389] | 0.580 [0.107] |
| **Model FT2** | 0.515 [0.291] | 0.394 [0.049] | **0.826 [0.084]** | 0.299 [0.368] | 0.704 [0.213] | 0.531 [0.342] |
| **Model FT3** | 0.738 [0.133] | 0.256 [0.127] | 0 [0.182] | **0.920 [0.040]** | 0.647 [0.238] | 0.648 [0.621] |
| **Model FT4** | 0.648 [0.178] | 0.519 [0.097] | 0.740 [0.094] | 0.538 [0.229] | 0.750 [0.136] | 0.650 [0.203] |
| **Model FTall** | 0.654 [0.137] | 0.344 [0.217] | 0.789 [0.072] | 0.542 [0.252] | 0.758 [0.130] | 0.662 [0.215] |

*All values are presented in the form Median [IQR], where IQR is the Interquartile Range between the third and the first quartile.*
*Legend: Verona = Multiple Sclerosis Centre of the University of Verona, MS_Lesion = ISBI 2015 Longitudinal MS Lesion Segmentation Challenge, HCP-A= Lifespan 2.0 Release of Human Connectome Project in Aging, Cam-CAN = Cambridge Centre for Ageing and Neuroscience, HCP-YA = Human Connectome Project Young Adult, General = General dataset composed of ten selected subjects for each of the five datasets (see Materials & Methods).*

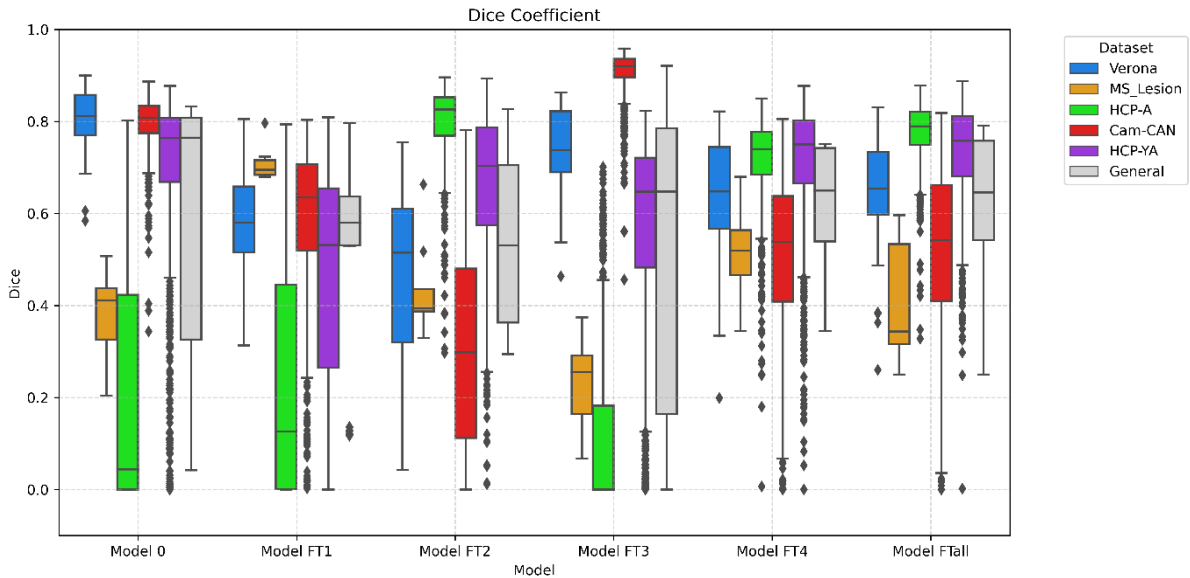

**Figure 2:** *Dice Coefficient values on the test sets of the five datasets and the General dataset are represented in the form of boxplots. The reference is the manual segmentation. Model 0 refers to the initial ASCHOPLEX trained on the Verona dataset, while Models FT1–4 correspond to the fine-tuning performed individually on MS_Lesion (Model FT1), HCP-A (Model FT2), Cam-CAN (Model FT3), and HCP-YA (Model FT4). Model FTall corresponds to the unique fine-tuning performed on all four datasets. General dataset is composed of ten selected subjects for each of the five datasets (see Materials & Methods).*

*ASCHOPLEX incremental learning with Dafne*

**Table 3** reports the results of the Dice Coefficient values on the testing subjects of all five investigated datasets comparing Model 0 and Model IL1-4. Unlike the previous approach, in the incremental learning step the sequential inclusion of additional datasets in the training phase does not negatively affect performance on the first dataset, Verona, which maintains Dice values in the median range 0.792-0.817. When considering each dataset individually, the best-performing model is not necessarily the one obtained after the incremental learning step on that specific dataset, as observed for HCP-A (Model IL2: 0.445 [0.438]; Model IL4: 0.534 [0.315]) or MS_Lesion (Model IL1: 0.498 [0.140]; Model IL2: 0.513 [0.161]). Nevertheless, training on a given dataset generally leads to improved Dice values on that dataset. Finally, when considering the General dataset, performance improves from Model 0 (0.765 [0.482]) to Model IL4 (0.788 [0.264]), indicating that incremental learning enhances cross-dataset generalization. In particular, Model IL4 achieves the highest median value with reduced variability both on General dataset and most of the single datasets, suggesting robust and consistent segmentation performance across diverse sources.

**Figure 3** shows the distributions of the Dice metrics for each model across the five datasets. Consistent with results in **Table 3**, from Model 0 to Model IL4, the distributions for Verona remain stable, with comparable median Dice and variability. In addition, HCP-A exhibits a more favorable distribution shape when the training reaches Model IL4. Finally, the distributions of the General dataset reflect the increasing generalizability achieved through the incremental learning phase, with higher median Dice value and lower variability for Model IL4 compared to the other models.

**Table 3**: *Dice Coefficient values on the test sets of the five datasets. The reference is the manual segmentation. Model 0 refers to the initial ASCHOPLEX trained on the Verona dataset, while Models IL1–4 correspond to incremental learning steps performed sequentially on MS_Lesion (Model IL1), HCP-A (Model IL2), Cam-CAN (Model IL3), and HCP-YA (Model IL4). General dataset is composed of ten selected subjects for each of the five datasets (see Materials & Methods).*

| Dice Coefficient | Verona | MS_Lesion | HCP-A | Cam-CAN | HCP-YA | General |
|---|---|---|---|---|---|---|
| **Model 0** | 0.811 [0.087] | 0.411 [0.111] | 0.044 [0.423] | 0.808 [0.059] | 0.764 [0.139] | 0.765 [0.482] |
| **Model IL1** | 0.817 [0.079] | 0.498 [0.140] | 0.112 [0.442] | 0.803 [0.074] | 0.719 [0.221] | 0.719 [0.404] |
| **Model IL2** | 0.817 [0.081] | 0.513 [0.161] | 0.445 [0.438] | 0.795 [0.083] | 0.753 [0.144] | 0.753 [0.344] |
| **Model IL3** | 0.806 [0.100] | 0.440 [0.163] | 0.299 [0.468] | 0.821 [0.064] | 0.721 [0.148] | 0.722 [0.501] |
| **Model IL4** | 0.792 [0.107] | 0.407 [0.119] | 0.534 [0.315] | 0.798 [0.068] | 0.788 [0.078] | 0.788 [0.264] |

*All values are presented in the form Median [IQR], where IQR is the Interquartile Range between the third and the first quartile.*

*Legend: Verona = Multiple Sclerosis Centre of the University of Verona, MS_Lesion = ISBI 2015 Longitudinal MS Lesion Segmentation Challenge, HCP-A= Lifespan 2.0 Release of Human Connectome Project in Aging, Cam-CAN = Cambridge Centre for Ageing and Neuroscience, HCP-YA = Human Connectome Project Young Adult, General = General dataset composed of ten selected subjects for each of the five datasets (see Materials & Methods).*

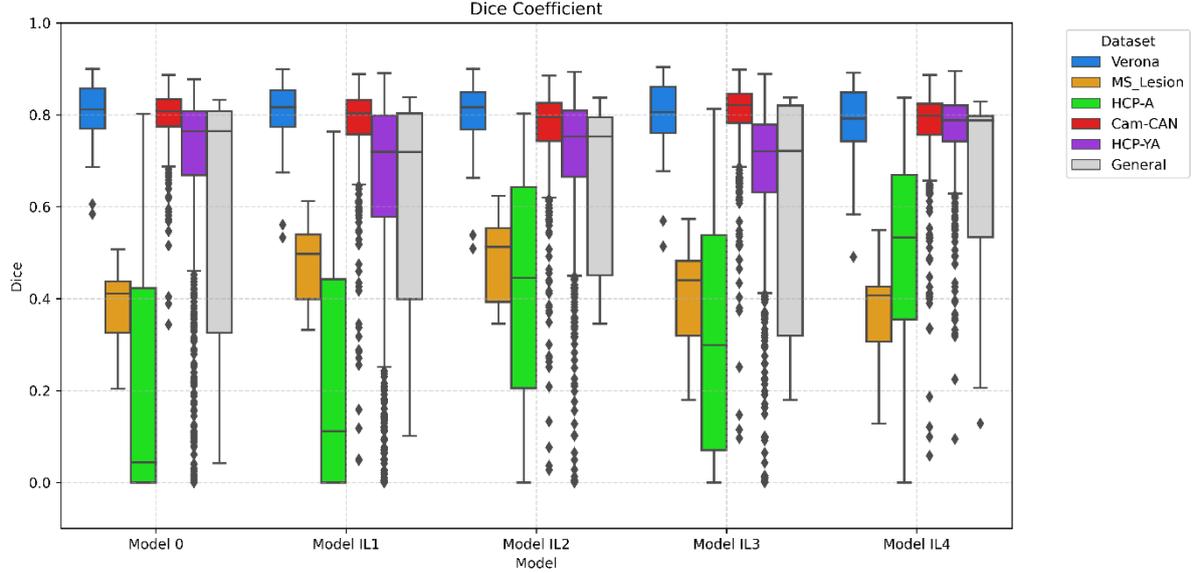

**Figure 3**: *Dice Coefficient values on the test sets of the five datasets and the General dataset are represented in the form of boxplots. The reference is the manual segmentation. Model 0 refers to the initial ASCHOPLEX trained on the Verona dataset, while Models IL1–4 correspond to incremental learning steps performed sequentially on MS_Lesion (Model IL1), HCP-A (Model IL2), Cam-CAN (Model IL3), and HCP-YA (Model IL4). General dataset is composed of ten selected subjects for each of the five datasets (see Materials & Methods).*

*Comparison of ASCHOPLEX fine-tuning and ASCHOPLEX incremental learning*

The Dice values reported in **Table 2** and **Table 3** show that Model FTX achieves higher performance on its corresponding dataset X than the respective Model ILX (i.e., MS_Lesion: Model FT1 0.695 [0.032], Model IL1 0.498 [0.140]; HCP-A: Model FT2 0.826 [0.084], Model IL2 0.445 [0.438]). However, the generalizability of Model ILX across all other datasets is higher than that of Model FTX, as reflected by the performance on the General dataset (e.g., Model FT1 0.580 [0.107] vs Model IL1 0.719 [0.404]; Model FT4 0.650 [0.203] vs Model IL4 0.788 [0.264]). Among the fine-tuned models, Model FTall achieves higher performance across datasets, and is therefore preferable to the individual Model FTX variants in terms of generalization, similarly to Model IL4, which represents the final stage of the incremental learning process. Based on these results, and with the aim of simulating a multi-center study, a comparison of generalizability between Model FTall and Model IL4 was conducted. Overall, Model IL4 achieves higher performance than Model FTall across all datasets (General – Model IL4: 0.788 [0.264]; Model FTall: 0.662 [0.215]) with the notable exception of HCP-A (Model IL4: 0.534 [0.315]; Model FTall: 0.789 [0.072]). These findings are corroborated by the scatterplots in **Figure 4**. When comparing the GT and the predicted ChPV, the regression line for Model IL4 is more parallel to the bisector of the first and third quadrants than that of Model FTall, except for HCP-A.

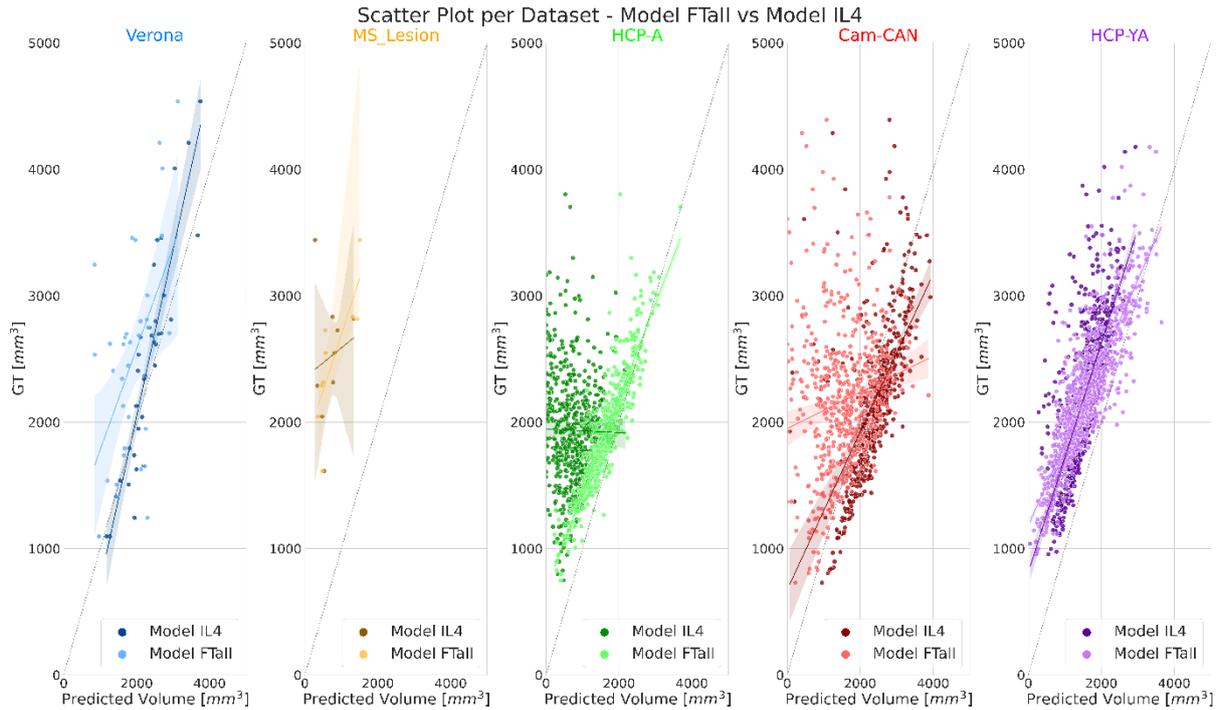

**Figure 4:** *Scatter plot and linear regression model fit with 95% confidence interval bounds of the Absolute Volume of the predicted segmentation (Predicted Volume) obtained on the testing subjects of the five datasets (Verona, MS_Lesion, HCP-A, Cam-CAN, HCP-YA) with the compared methods (Model FTall, Model IL4), and manual segmentation (GT), used as reference.*

**Discussion:**

In this study, we introduced an enhanced version of ASCHOPLEX (Visani et al., 2024b). This framework outperforms state-of-the-art methods for the automatic and reliable segmentation of the human ChP from T1-w brain MRI. Despite its strong performance, the original fine-tuning approach has several limitations in specific contexts. These are mainly related to the limited variability of the original training dataset, which restricts the model's generalizability to previously unseen datasets, thereby necessitating a fine-tuning step. However, fine-tuned models are susceptible to the issue of catastrophic forgetting, as previously learned knowledge may be lost during adaptation to new data (Aleixo et al., 2023). A straightforward solution to this problem would be to pool data across multiple datasets, but data protection regulations often impede data sharing across institutions (Raudys and Jain, 1991), further complicating model adaptation. An additional constraint of the fine-tuning approach is its reliance on annotated data, as the method is fully supervised. In the case of multi-scanner datasets, separate fine-tuned models are often required to ensure optimal performance.

To address these limitations, this work investigates the generalizability of ASCHOPLEX by comparing its original fine-tuning approach with a novel incremental learning strategy integrated into the Dafne framework (Santini et al., 2025). The comparison is conducted across five distinct datasets, with the aim of identifying optimal strategies for enhancing model robustness and adaptability in diverse imaging contexts. Specifically, we compared the starting model obtained from the initial training on the reference dataset (Model 0), the single fine-tuned models (Model FTX), the model derived from the fine-tuning on all evaluated datasets (Model FTall), and the models derived from the incremental learning procedure in Dafne (Model ILX).

*The new version of ASCHOPLEX*

Data collections were selected to maximize variability across several dimensions, including manufacturer, scanner type, acquisition sequence, spatial resolution, and overall image quality (e.g., presence of noise and artifacts). Publicly available datasets such as HCP-A, HCP-YA, and Cam-CAN were chosen to simulate a realistic large-scale population. The preprocessing step led to the exclusion of a large number of subjects from these datasets due to motion-related artifacts - particularly in HCP-A - highlighting the inherent variability within these data sources.

The proposed updated version of ASCHOPLEX addresses the challenge of segmenting the ChP in low-quality images by incorporating additional data augmentation techniques and the SwinUNETR architecture (Hatamizadeh et al., 2022a). The selection of the three architectures - UNETR, DynUnet, and SwinUNETR - as the best performing models across the five training folds demonstrates the added value contributed by swin transformers to the ChPV segmentation task.

*ASCHOPLEX fine-tuning or ASCHOPLEX incremental learning?*

The fine-tuning of ASCHOPLEX demonstrates its capacity to effectively learn and adapt to new dataset-specific characteristics. This is evidenced by a marked increase in the Dice Coefficient from the baseline model (Model 0) to the corresponding fine-tuned models (Models FT1–4) across each target dataset. However, the fine-tuned models experience catastrophic forgetting when re-tested on the original dataset or evaluated on unseen data. Fine-tuning on all four datasets (Model FTall) yields suboptimal results on individual datasets compared with the corresponding single fine-tuned models due to the high data variability and the limited number of subjects included in the training step. Nevertheless, the generalizability is higher as the model learns from highly variable data, although catastrophic forgetting remains present. These findings suggest that ASCHOPLEX fine-tuning is particularly advantageous when the target dataset is homogeneous and standardized (i.e., images are always acquired with the same MRI scanner and acquisition protocol), whereas it is less efficient and the performance deteriorates when data variability is high, as in multi-center studies.

When the study is multi-center or the dataset exhibits high variability (i.e., image acquisition protocol or MRI scanner change), ASCHOPLEX incremental learning with the Dafne platform offers a solution to the limitations of the fine-tuning approach. In fact, the incremental learning strategy enables the assimilation of new features while preserving previously acquired knowledge, thereby mitigating the issue of catastrophic forgetting. From Model 0 to Model IL4, the median Dice Coefficient on the cumulative dataset remains consistently stable, with the highest performance observed in the model that has integrated information from all datasets. Notably, no substantial performance degradation is observed across any individual dataset throughout the incremental learning process, particularly re-testing on the original dataset (Verona). Although the incremental learning approach does not always reach the same level of performance on newly introduced datasets as the single fine-tuned model, its overall performance trajectory is more stable and uniformly increasing, thus addressing the generalizability requirement across different scenarios.

Additional advantages of ASCHOPLEX integrated within Dafne compared with its fine-tuned version are as follows. Both ASCHOPLEX implementations are freely available (see *Software Availability*), but they differ in user-friendliness. The Dafne framework provides a graphical user interface - absent in the fine-tuning version - and is easier to use than the Python-based pipeline for non-experts. Moreover, the challenge of requiring previously annotated data in the standard ASCHOPLEX version is mitigated in Dafne, which enables inference using the most recently updated model on the training images. The resulting segmentations can be manually corrected and saved, significantly reducing the time and effort required for manual annotation, particularly when working with entirely unlabeled datasets. Finally, the integrated federated learning framework allows model weights to be shared without requiring access to the original data, thereby preserving data privacy and eliminating the need to centralize all data for training a single, unified model (such as Model FTall).

*Limitations*

This study has several limitations. First, all datasets consist of images acquired exclusively using 3T MRI scanners. Second, the analysis was restricted to scanners from only two manufacturers (Philips and Siemens), which may limit the assessment of generalizability across other vendors. Third, the investigated populations included only HC and MS, thereby excluding other potential clinical conditions, and the datasets differ in terms of clinical status (HC vs. MS), age range, sample size, and image quality. Fourth, the stability of the models in the incremental learning approach (Models IL1–4) when changing the order of the datasets was not assessed in this study to simulate a realistic multi-center scenario. Fifth, the number of subjects used for fine-tuning was set to ten (five for training and five for validation), following previous analyses (Visani et al., 2024b). Lastly, the number of subjects used for incremental learning in Dafne was also set to ten to ensure consistency with the fine-tuning approach, while the model weights for incremental learning were kept at the Dafne default values. Future studies will examine the influence of these parameters on model performance.

**Conclusion:**

ASCHOPLEX is an effective tool for the segmentation of the ChPV. However, the generalizability provided by its fine-tuning approach is limited, particularly when applied to heterogeneous datasets. In such scenarios, the incremental learning version of ASCHOPLEX - integrated within the Dafne federated learning framework - offers a viable alternative. Beyond enhancing generalizability, this strategy mitigates catastrophic forgetting, thereby preserving previously acquired knowledge, ensuring superior robustness, consistently high accuracy, and stable performance across diverse datasets.

**Acknowledgements:**

We would like to thank the Cam-CAN, HCP-A, and HCP-YA imaging initiatives, whose support was indispensable for this study.
Cam-CAN funding was provided by the UK Biotechnology and Biological Sciences Research Council (grant number BB/H008217/1), with additional support from the UK Medical Research Council and the University of Cambridge, UK.
The HCP-A project was supported by the National Institute on Aging of the National Institutes of Health under Award Number U01AG052564, as well as by funding from the McDonnell Center for Systems Neuroscience at Washington University in St. Louis. The HCP-A-Aging 2.0 Release data used in this study is available at DOI: 10.15154/1520707.
The HCP-YA Data were provided (in part) by the Human Connectome Project, WU-Minn Consortium (Principal Investigators: David Van Essen and Kamil Ugurbil; 1U54MH091657) funded by the 16 NIH Institutes and Centers that support the NIH Blueprint for Neuroscience Research; and by the McDonnell Center for Systems Neuroscience at Washington University.